\documentclass[journal]{IEEEtran}
\usepackage{graphicx,amssymb,cite,amsbsy,subfigure,amsmath,amsfonts,multirow,color,array,algorithm,algorithmic,lineno, enumerate, bm,setspace, booktabs, multirow}
\def\0{\mathbf{0}}
\def\1{\mathbf{1}}
\def\1{\mathbf{2}}
\def\1{\mathbf{3}}
\def\1{\mathbf{4}}
\def\1{\mathbf{5}}
\def\1{\mathbf{6}}
\def\1{\mathbf{7}}
\def\1{\mathbf{8}}
\def\1{\mathbf{9}}

\begin{document}
	%
	\title{Multi-Attention Generative Adversarial Network for Remote Sensing Image Super-Resolution}
	%
	%
	%
	\author{Meng Xu, Zhihao Wang, Jiasong Zhu, Xiuping Jia, Sen Jia}
	
	\maketitle
	
	\begin{abstract}
		
		Image super-resolution (SR) methods can generate remote sensing images with high spatial resolution without increasing the cost, thereby providing a feasible way to acquire high-resolution remote sensing images, which are difficult to obtain due to the high cost of acquisition equipment and complex weather. Clearly, image super-resolution is a severe ill-posed problem. Fortunately, with the development of deep learning, the powerful fitting ability of deep neural networks has solved this problem to some extent. In this paper, we propose a network based on the generative adversarial network (GAN) to generate high resolution remote sensing images, named the multi-attention generative adversarial network (MA-GAN). We first designed a GAN-based framework for the image SR task. The core to accomplishing the SR task is the image generator with post-upsampling that we designed. The main body of the generator contains two blocks; one is the pyramidal convolution in the residual-dense block (PCRDB), and the other is the attention-based upsample (AUP) block. The attentioned pyramidal convolution (AttPConv) in the PCRDB block is a module that combines multi-scale convolution and channel attention to automatically learn and adjust the scaling of the residuals for better results. The AUP block is a module that combines pixel attention (PA) to perform arbitrary multiples of upsampling. These two blocks work together to help generate better quality images. For the loss function, we design a loss function based on pixel loss and introduce both adversarial loss and feature loss to guide the generator learning. We have compared our method with several state-of-the-art methods on a remote sensing scene image dataset, and the experimental results consistently demonstrate the effectiveness of the proposed MA-GAN.
			
	\end{abstract}\label{abstract}

	\begin{IEEEkeywords}
		Super-resolution (SR), Multi-attention, Generative adversarial nets (GANs), Remote sensing image
	\end{IEEEkeywords}
	
	\IEEEpeerreviewmaketitle
	
	\section{Introduction} \label{sec:intro}
	
	\begin{figure*}[h]
		\centering
		\includegraphics[width=0.9\linewidth]{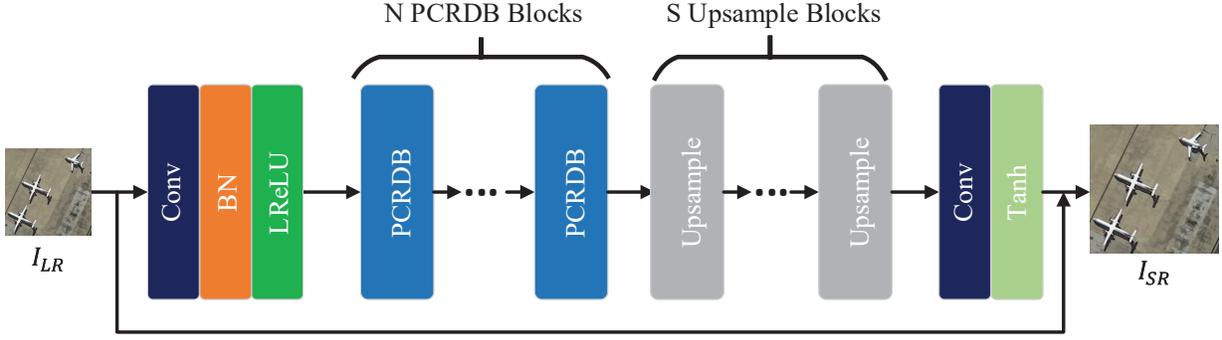}
		\caption{Architecture of the MA-GAN generator.}
		\label{fig:generator}
	\end{figure*}	
	\begin{figure*}[h]
		\centering
		\includegraphics[width=0.9\linewidth]{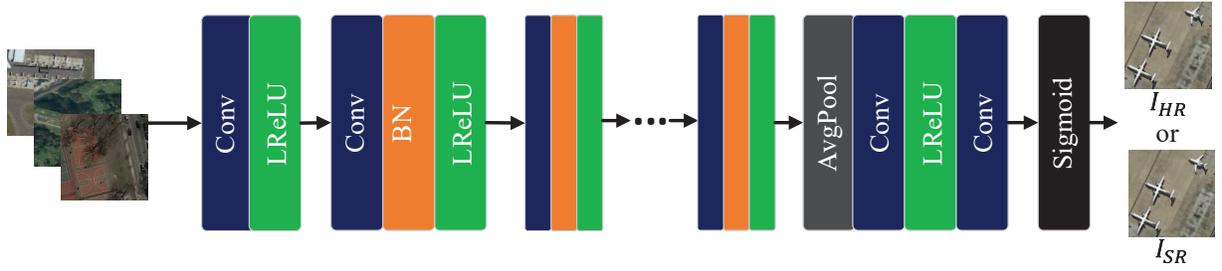}
		\caption{Architecture of the MA-GAN discriminator.}
		\label{fig:discriminator}
	\end{figure*}
	
	Image super-resolution (SR) is a rapidly growing issue in computer vision and has received considerable critical attention in recent years. Image SR aims to recover high-resolution (HR) images from given low-resolution (LR) images. It is essential for a wide range of real-world applications such as medical image processing \cite{isaac2015srformedicalimg, greenspan2009srinmedical, huang2017simultaneoussr}, compressed image/video enhancement \cite{li2016videosradaptive, caballero2017videosr}, surveillance and security \cite{rasti2016facesurveillance, zou2011lrface}, etc. Image super-resolution is not only about improving the visual quality of images: more importantly, almost all vision tasks can benefit from it. HR images generated by SR methods offer more options for vision tasks because they contain richer information.
	
	As for the field of remote sensing, image super-resolution is even more significant. SR methods can facilitate other remote sensing missions, such as target detection \cite{bai2018sod, zou2016ship, tatem2001srtarget}, environmental monitoring \cite{cheng2020reslap, elfadaly2018monitoring, liu2017change}, scene analysis \cite{shi2017descrsi} and so on. It is common knowledge that remote sensing images are generally acquired by satellites from high altitude. This results in the low spatial resolution of remote sensing images. The quality of remote sensing images is also affected by many factors, such as motion blur, atmospheric interference (e.g., cloud cover), ultra-long-range imaging, transmission noise, etc. \cite{yu2020edbpn, zhang2020mhan}. Some of these problems can be solved by utilizing more advanced equipment, but this equipment is more expensive with respect to both launch deployment and routine maintenance. It is therefore easy to conclude that the most appropriate way to obtain high-resolution images is through SR algorithms.
	
	Unfortunately, image super-resolution is a severely ill-posed problem, as the process of image degradation is not unique and there is therefore no certain solution. Traditional image super-resolution algorithms are basically interpolation-based or reconstruction-based. Interpolation-based methods obtain a pixel value by using its neighborhood information, and interpolation algorithms such as linear and bicubic \cite{keys1981cubic} interpolation and lanczos \cite{duchon1979lanczos} upsampling are essentially different with respect to neighborhood selection and calculation methods. Although interpolation-based methods are less computationally demanding, they often fail to recover the high-frequency information and therefore result in an excessively smooth image with blurred details, especially edges and texture information. For reconstruction-based SR algorithms, they achieve better results by using different prior knowledge or constraints in the form of distribution or energy function between HR and LR images. Local \cite{sun2008srgrad, tai2010sredge}, nonlocal \cite{yang2010exploiting} and sparse priors \cite{dong2016hyperspectralsr, lu2013imagesr} are the most widely used constraints for image super-resolution tasks. Reconstruction-based methods mostly use singular prior knowledge or a combination of several instances of prior knowledge. However, in comparison to methods based on interpolation, they are more computationally demanding. Meanwhile, these manually designed priors may not perform well when the scene changes.
	
	Recently, with the development of machine learning (ML), learning-based SR algorithms have emerged. These algorithms attempt to establish an implicit mapping from LR images to the corresponding HR images through ML models. Among these algorithms, there has been a huge explosion in the number of deep learning (DL)-based SR algorithms. In this paper, we only focus on DL-based SR algorithms. Since Dong et al. proposed SRCNN \cite{dong2014srcnn} in 2014, an increasing number of studies have been performed by using convolutional neural networks (CNNs) for SR tasks, due to the powerful non-linear fitting and learning capabilities of CNNs. At the same time, with the development of several excellent CNNs (ResNet \cite{he2016resnet}, DenseNet \cite{huang2017densenet}, etc.), more CNN-based SR networks have emerged, such as VDSR \cite{kim2016vdsr}, EDSR \cite{lim2017edsr}, DBPN \cite{haris2018dbpn} and others. These end-to-end models need to be trained with a large number of paired LR and HR images to learn the mapping from LR to HR images. Thanks to the advancement of DL, these models can use deeper and more complex networks to learn higher-level features and thus produce higher-quality HR images. Although CNN-based SR algorithms outperform the others, remote sensing images are somehow different from natural images. Compared with natural images, the target in a remote sensing image covers fewer pixels and the background is more complex. For example, a desert image may contain less textures, but a dense residential area will contain an extremely rich amount of textural information, making super-resolution of remote sensing images more difficult. The current work \cite{yu2020edbpn, zhang2020sceneadapt} already addresses some of these issues to some extent.
	
\begin{figure*}[t]
		\centering
		\includegraphics[width=0.9\linewidth]{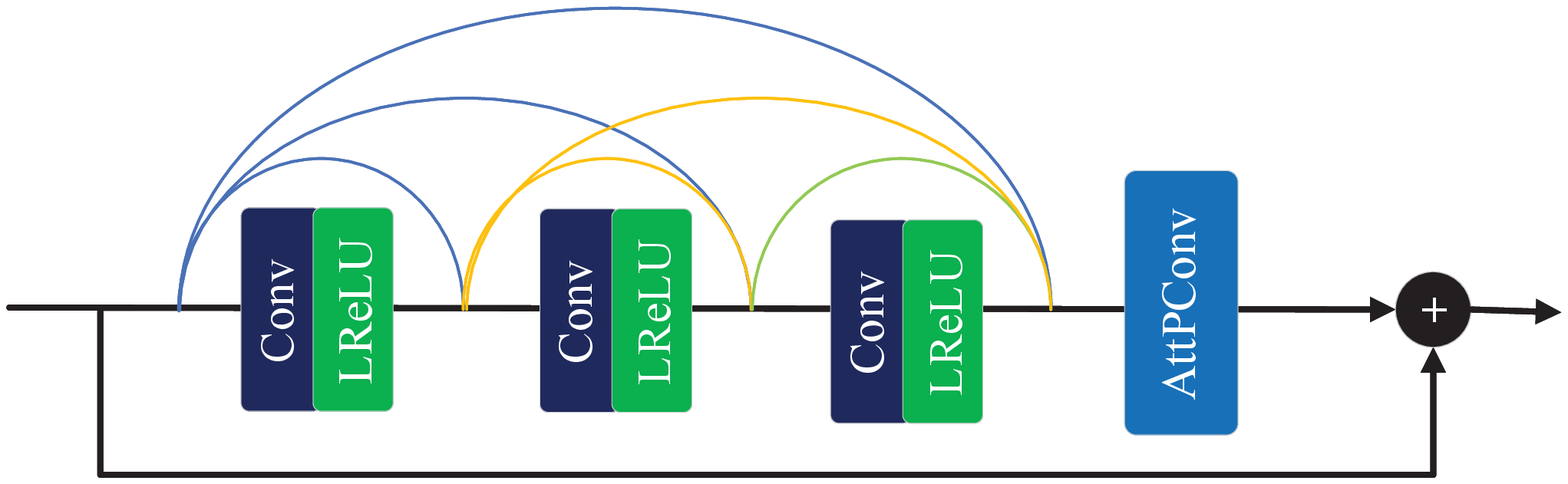}
		\caption{The PCRBD Block.}
		\label{fig:PCRDB}
	\end{figure*}

\begin{figure*}[]
		\centering
		\subfigure[The AttPConv Block.]{\label{subfig:AttPConv}
			\includegraphics[width=0.9\columnwidth]{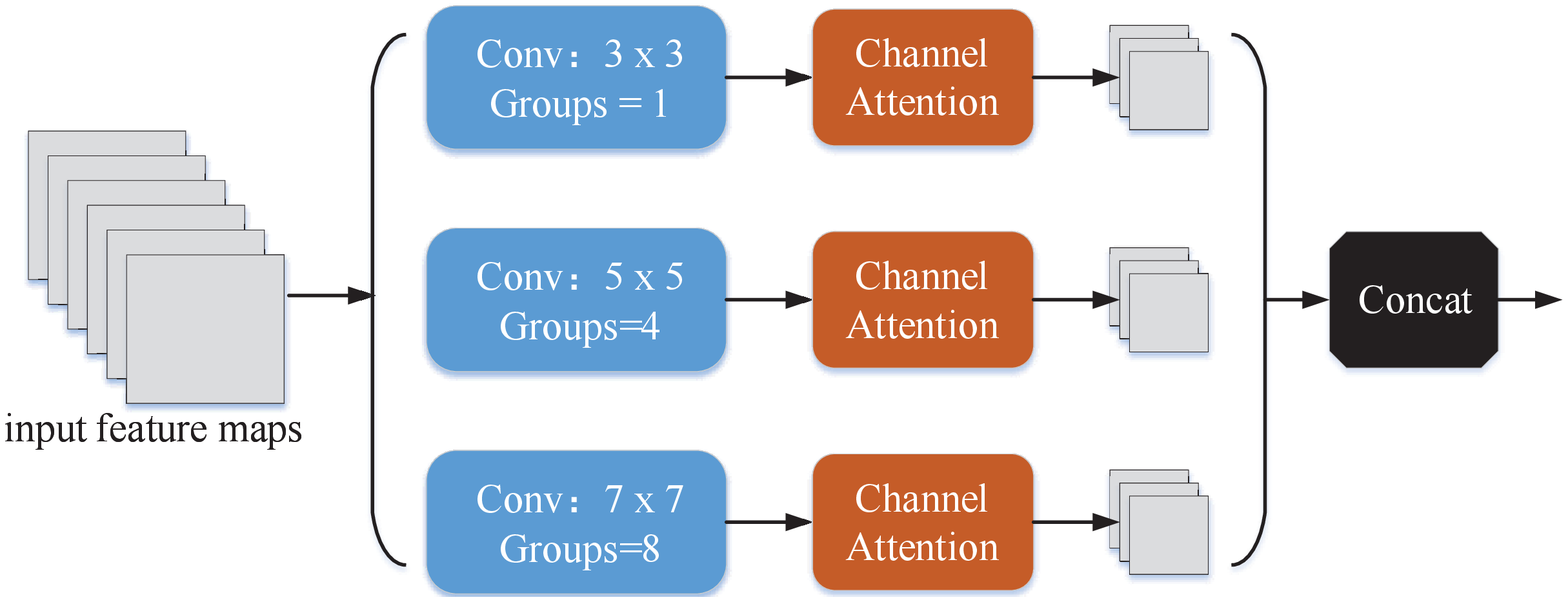}
		}
		\subfigure[The AUP Block.]{\label{subfig:Upsample}
			\includegraphics[width=0.9\columnwidth]{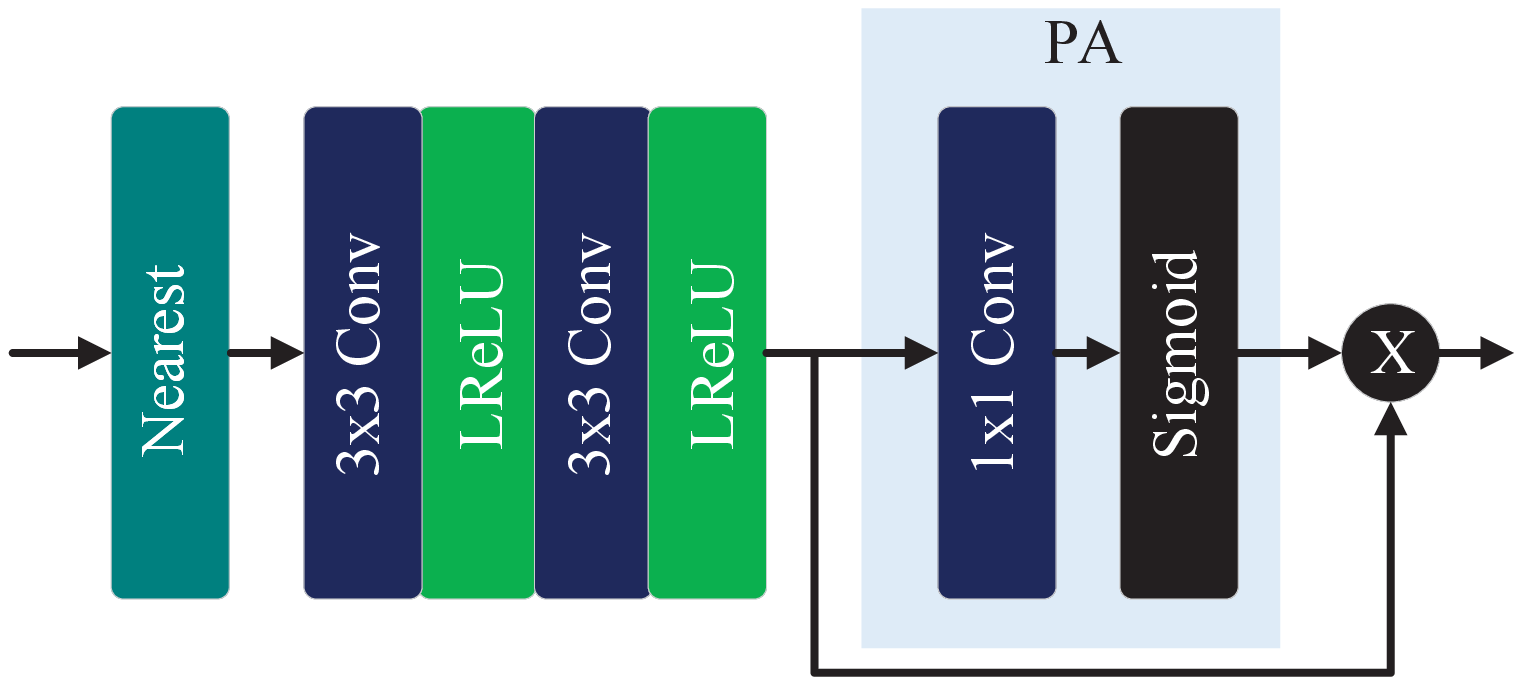}	
		}
		\caption{Modules containing attention mechanisms}
	\end{figure*}

	In this work, we proposed a GAN \cite{goodfellow2014gan}-based SR network that increases the resolution for remote sensing images in a satisfactory way, which is called the multi-attention super-resolution network (MA-GAN), as shown in Figure \ref{fig:generator} and Figure \ref{fig:discriminator}. The main body of the generator, which is the core of the super-resolution task, consists of two modules; one is the PCRDB block (Figure \ref{fig:PCRDB}), and the other is the AUP block (Figure \ref{subfig:Upsample}).
The structure of the PCRBD block is similar to that of ESRGAN \cite{wang2018esrgan}, which includes skip \cite{he2016resnet} and dense \cite{huang2017densenet} connections to better extract features. However, we replace the last convolution operation with an AttPConv of our own design. The AttPConv block first performs multi-scale convolution without requiring more computations, and subsequently computes channel attention \cite{zhang2018rcan} on the feature maps obtained after convolution at different scales in order to dynamically adjust each feature map to achieve better SR performance. Compared with ESRGAN, which uses a fixed parameter (which needs to be designed manually) to scale the residuals, our PCRBD undoubtedly makes more full use of the powerful learning ability of CNNs and is more generalizable. The AUP block is a module that can upsample the input feature map by an arbitrary multiple. It first upsamples the input by nearest neighbor interpolation and then fine-tunes the result of the upsampling by means of a small neural network and PA \cite{zhao2020pan}. Since simple upsampling can lead to excessively smooth output, the design of the AUP block can compensate for the drawbacks caused by interpolation and thus enable better SR results.
	
	In summary, the contributions of this work are as follows:
	
	\begin{enumerate}[1.]\setlength{\itemindent}{1em}
		\item First, we proposed a GAN-based SR network that offers the ability of increasing the resolution for remote sensing images with any scale factor. Our experiments on the NWPU-RESISC45 dataset demonstrate that the performance of the proposed MA-GAN approach outperforms the state-of-the-art SR methods.
		\item We design the PCRBD block that contains AttPConv. PCRDB extracts features better by skip and dense connections, and features at different scales can be obtained by multi-scale convolution in the final AttPConv block, and then the residuals can be adjusted for each feature map by the final channel attention.
		\item We design the AUP block. With its nearest neighbor interpolation at the beginning, the AUP can achieve upsampling at any scale. The small CNN and PA immediately after interpolation can be adjusted based on the interpolated feature maps to achieve better upsampling results.
	\end{enumerate} 
	
	The remainder of this paper is organized as follows. Section \ref{sec:related} briefly reviews the related works. Section \ref{sec:method} presents the proposed MA-GAN in detail. Experimental results are shown in Section \ref{sec:experiment}. Finally, the conclusion is presented in Section \ref{sec:conclusion}.
	

	\section{Related Work} \label{sec:related}
	
	\subsection{Upsampling Method} \label{subsec:up_methods}	
	The aim of image SR is to improve the resolution of the image, so the upsampling method is indispensable for any SR algorithm, and the way in which it is performed will also affect the final result. There are three commonly used upsampling methods. The first one is interpolation. The second is the subpixel convolution layer \cite{shi2016espcn} proposed by Shi et al., which transforms the feature map $F_a \in \mathbb{R}^{h \times w \times r^2c}$ into a new feature map $F_b\in \mathbb{R}^{rh \times rw \times c}$ by a scale factor of $r$ ($h, w$ and $c$ are the height, width and channels of feature maps). The last upsampling method is transposed convolution, also called deconvolution in some papers. The most common approach to using transposed convolution is the up- and down-projection unit module proposed in the DBPN \cite{haris2018dbpn}. In recent studies, researchers have favored the direct use of interpolation for upsampling. This is because operations such as transposed convolution can introduce checkerboard-like artifacts that affect the experimental results, and subpixel and transposed convolution can only upsample integer multiples, which is more restrictive. In this paper, interpolated upsampling is used to achieve super-resolution at any scale factor.

	\subsection{Super-Resolution Framework} \label{subsec:framework}
	SR networks can be divided into four types, depending on where the upsampling operation is located in the network: pre-upsampling, post-upsampling, progressive upsampling, and iterative up and down sampling SR frameworks \cite{haris2018dbpn}. The pre-upsampling framework \cite{dong2014srcnn, dong2015srcnn} achieves the difficult upsampling task by interpolation at the beginning and later refines the interpolated image. The pre-upsampling framework is easier to train because only the coarsest images need to be refined, but it costs more time and space because of its higher-dimensional feature maps. The post-upsampling framework \cite{ledig2017srgan, lim2017edsr} can be trained much more rapidly because it includes the upsampling step at the end. By dividing the upsampling into smaller tasks, the progressive upsampling framework \cite{lai2017lapsrn} not only speeds up training but also yields better results. The iterative up and down sampling framework \cite{haris2018dbpn, yu2020edbpn} allows for better mining of the relationships between paired LR and HR images, but because of its complex framework and large number of up- and down-sampling operations, it often requires heavy manual network design.

	\subsection{Loss Function} \label{subsec:r_loss}
	The loss function is one of the most essential aspects in image SR tasks since one LR image could correspond to several HR images. A suitable loss function brings the generative model closer to the true HR image area in its latent space.

	We denote $I_{LR}$ for an LR image, $I_{HR}$ for an HR image and $I_{SR}$ for the generated HR image. Almost all SR algorithms introduce a pixelwise loss that brings $I_{SR}$ closer to $I_{HR}$ in pixel value.	$L_1$ loss and $L_2$ loss are the most frequently used pixelwise losses. The pixel-level loss improves the peak signal-to-noise ratio (PSNR) of $I_{SR}$, but results in loss of high-frequency information. Therefore, Lefig et al. \cite{ledig2017srgan} presented perceptual loss with respect to perceptually relevant characteristics. GAN-based models have adversarial loss, which helps the generator to produce higher quality images with the help of a discriminator. In addition to these losses, there are also texture losses \cite{gatys2015texture}, cycle consistency loss \cite{yuan2018unsupervised}, etc.

	\section{Proposed Multi-attention Generative Adversarial Nets} \label{sec:method}

	In this section, we will introduce the proposed MA-GAN in detail. First, we will present the overall framework of the GAN used in this paper. Second, we will introduce the multi-attention mechanism and describe each block in detail. Finally, we will present a description of the loss function used in this paper.
	
	\subsection{Network Framework} \label{subsec:arch}
	
	To be able to generate higher quality HR images, we design a multi-attention generative adversarial network. It contains a generator and discriminator as shown in Figure \ref{fig:generator} and Figure \ref{fig:discriminator}, respectively.
	
	To speed up the training, we design a generator which belongs to a post-upsampling framework. The generator network can be divided into four parts. The head of the generator is a Convolution-BatchNormalize-LeakyReLU block, while the body part consists of $N$ PCRDB modules, followed by $S$ upsampling blocks. Finally, the tail of the generator consists of the convolution-Tanh block. In this paper, we consider the SR task as an optimization task, so we perform pixel-level summation of the output of the tail with interpolated $I_{LR}$, which has the same spatial resolution of $I_{HR}$. Since all GANs are difficult to train, this operation reduces the difficulty of training and shortens the time consumption to obtain the desired model. The detailed architecture of the PCRBD and AUP blocks will be introduced in Section \ref{subsec:PCRDB} and Section \ref{subsec:up_block}.
	
	The discriminator we use is the same as SRGAN \cite{ledig2017srgan}, which mainly consists of a Convolution-BatchNormalize-LeakyReLU block. The generated image $I_{SR}$ is fed into the discriminator together with $I_{HR}$ to calculate adversarial loss to guide the generator training.

	\subsection{Pyramidal Convolution in Residual-Dense Block}\label{subsec:PCRDB}
	
		To introduce PCRBD, we first present the AttPConv block. AttPConv is a special convolution combining multi-scale convolution and multi-channel attention. As shown in Figure \ref{subfig:AttPConv}, our proposed multi-scale convolution is not simply composed of multiple parallel conventional convolutions, and we prefer to call it pyramidal convolution. The pyramidal convolution is implemented internally by grouping convolution without increasing the computational cost or the model complexity \cite{duta2020pyconv}. The pyramid convolution in this paper has three different kernels, $3\times 3$, $5\times 5$ and $7\times 7$, corresponding to the feature maps groups of 1, 4 and 8, respectively. Each group convolution is followed by a channel attention module \cite{zhang2018rcan}, and all the feature maps are concatenated together as the final output. Based on this structure, the AttPConv block has the same input and output forms as the conventional convolution.

		Since the AttPConv module contains an attention module, it is used as the last convolutional module of the PCRDB module. AttPConv follows 3 more Convolution-LeakyReLU modules, as shown in Figure \ref{fig:PCRDB}. They are connected by dense connection \cite{huang2017densenet}. The final output feature maps are obtained by element-level addition of the output of AttPConv and the input for the PCRDB module. It is worth mentioning that the residual-in-residual dense block (RRDB) proposed in ESRGAN has a structure similar to PCRDB. However, the RRDB block uses residual scaling \cite{lim2017edsr, szegedy2017inceptionv4} to reduce the residuals by multiplying a constant between 0 and 1 before adding them to the main path to prevent instability. This constant can only be determined experimentally, which entails some additional work. We not only use convolution kernels of different size to extract feature information at different scales by the last AttPConv block in PCRDB. They also fulfill the role of scaling constant by channel attention, which is equivalent to multiplying a dynamic scaling constant, and this value is different for each feature map, thus scaling the residuals more accurately.

	\subsection{Attention-Based Upsample Block}\label{subsec:up_block}
	
		We mentioned several upsampling methods in Section \ref{subsec:up_methods}, and since the subpixel convolution layer \cite{shi2016espcn} and transposed convolution present some drawbacks, we use an interpolation-based upsampling method in this paper. However, simple interpolation can cause excessive smoothing of the image; thus, we design an upsampling module based on the pixel attention, as shown in Figure \ref{subfig:Upsample}.
		
		First, we perform nearest neighbor interpolation on the input of the upsampling module to improve the spatial resolution of the feature map. It is then fed into two Convolution-LeakyReLU blocks, and the feature map of the final Convolution-LeakyReLU block is denoted as $F_{up} \in \mathbb{R}^{h \times w \times c}$. The working process of the PA module can be described by the following Equation (\ref{pa}).
			
		\begin{equation}\label{pa}
			F_{PA}= Sigmoid\left (Con1\times1\left ( F_{up} \right )  \right )
		\end{equation}
		Here, $F_{PA}$ refers to the output of the PA module and $Con1\times1$ represents the convolution operation with a kernel size of $1\times1$. The channel number of the feature maps is reduced to $1$ after $Con1\times1$, and then the $F_{PA} \in \mathbb{R}^{h \times w \times 1}$ is obtained after sigmoid operation. The values of $F_{PA}$ are between 0 and 1 so that the feature map can be adjusted at the element level to produce better SR results.

	\subsection{Loss Function} \label{subsec:loss}
	
	In general, $L_2$ loss speeds up the training process, but it can introduce gradient explosion, and therefore the results are less robust. $L_1$ loss has a stable gradient and is more robust, but in contrast, the training is slower and the solution is less stable. Since $L_1$ and $L_2$ losses both exhibit their own advantages and disadvantages, we choose smooth-$L_1$ loss, which combines the advantages of $L_1$ and $L_2$ losses, as the pixelwise loss. The smooth-$L_1$ loss is calculated as:
	
	\begin{equation}\label{eq:smoothl1_1}
	l_{pixel}= \begin{cases}
	0.5 \times \frac{1}{r^{2}HWC}\sum x^{2}
	& \text{if} \; \left |x\right |< 1 \\
	\frac{1}{r^{2}HWC}\sum \left(\left | x \right | - 0.5\right)
	& \text{otherwise}
	\end{cases}
	\end{equation}
where $x =  I_{HR} - I_{SR}$. $H$, $W$, and $C$ are the height, width and channels of $I_{LR}$, respectively. 
	
	$l_{pixel}$ makes the $I_{SR}$ as close to the corresponding $I_{HR}$ as possible at each pixel value to obtain high PSNR. However, this tends to ignore the high frequency information of the image and yield perceptually unsatisfying images. To solve this, inspired by the perceptual loss in SRGAN \cite{ledig2017srgan}, we design the feature loss $l_{fea}$. The $l_{fea}$ is based on the input for the last average pooling layer of the pretrained ResNet-18 \cite{he2016resnet} model. We denote the process of obtaining the desired feature map by $Extractor(\cdot)$. The feature loss $l_{fea}$ is then defined as:
	
	\begin{equation}\label{eq:fea_loss}
		l_{fea} = \frac{1}{hwc}\sum {[Extractor(I_{HR})-Extractor(I_{SR})]}^{2}
	\end{equation}
	
	We also introduced $l_{adv}$ from the discriminator to assist in the training of the generator. The adversarial loss $l_{adv}$ for $G$ is:	
	\begin{equation}\label{eq:adv_loss}
	l_{adv} = -log\left ( D(I_{SR}) \right ) = -log\left ( D(G(I_{LR})) \right )
	\end{equation}	
	where $G\left(\cdot\right)$ denotes the output of generator and $D\left(\cdot\right)$ denotes the output of discriminator.
		
	The total loss for the generator is:
	\begin{equation}\label{eq:g_loss}
	l_{G} = l_{pixel} + \alpha l_{adv} + \beta l_{fea}
	\end{equation}
	where $\alpha$ and $\beta$ are the weighting parameters for $l_{adv}$ and $l_{fea}$, respectively.
	
	For the discriminator, we use a loss function consistent with other GANs, shown as follows:
	\begin{equation}\label{d_loss}
		l_D = -log(D(I_{HR}))-log(1-D(I_{SR}))
	\end{equation}

	\section{Experimental Results} \label{sec:experiment}
	
	\subsection{Dataset}\label{subsec:dataset}
	
	\begin{table}[]
		\renewcommand{\arraystretch}{1.1}
		\caption{The descriptors of different scenes.}
		\centering
		\label{tab:scene_no}
		\begin{tabular}
			{p{0.06\columnwidth}<{\centering}|p{0.35\columnwidth}<{\centering}|p{0.06\columnwidth}<{\centering}|p{0.35\columnwidth}<{\centering}}
			\hline
			\hline
			No. & Scene                & No. & Scene                 \\
			\hline
			S1  & parking lot          & S24 & baseball diamond      \\
			S2  & runway               & S25 & intersection          \\
			S3  & island               & S26 & meadow                \\
			S4  & sparse residential   & S27 & circular farmland     \\
			S5  & commercial area      & S28 & basketball court      \\
			S6  & terrace              & S29 & beach                 \\
			S7  & tennis court         & S30 & sea ice               \\
			S8  & industrial area      & S31 & roundabout            \\
			S9  & wetland              & S32 & dense residential     \\
			S10 & desert               & S33 & chaparral             \\
			S11 & stadium              & S34 & railway station       \\
			S12 & overpass             & S35 & thermal power station \\
			S13 & snow berg             & S36 & harbor                \\
			S14 & medium residential   & S37 & storage tank          \\
			S15 & freeway              & S38 & forest                \\
			S16 & ground track field   & S39 & railway               \\
			S17 & church               & S40 & airport               \\
			S18 & mobile home park     & S41 & palace                \\
			S19 & lake                 & S42 & ship                  \\
			S20 & rectangular farmland & S43 & golf course           \\
			S21 & mountain             & S44 & airplane              \\
			S22 & bridge               & S45 & cloud                 \\
			S23 & river                &     &  \\
			\hline
			\hline
		\end{tabular}
	\end{table}
	
	We perform experiments on the widely used remote sensing image dataset NWPU-RESISC45 \cite{cheng2017nwpu}. This dataset contains a total of 31,500 images with $256 \times 256$ pixels covering 45 scenes. NWPU-RESISC45 contains large-scale remote sensing images that vary greatly in translation, spatial resolution, viewpoint, object pose, illumination, background, and occlusion. It also has high within-class diversity and between-class similarity. Therefore, it is a very challenging dataset for image SR. In this paper, we randomly divide the images of each of the scenes in the dataset into a training set, a validation set and a test set at a ratio of $8:1:1$. All $I_{HR}$ are resized to $224 \times 224$ in the experiments.

	\subsection{Implementation Details} \label{subsec:implementation}
	
	We trained our network on a platform equipped with the Intel(R) Xeon(R) CPU (2.50 GHz) with 250-GB memory, an NVIDIA Tesla P100 with 16-GB memory and an AVAGO MR9361 with 2-TB capacity.
	
	Our experiments used three scale factors: the simpler upsampling by a factor of 2, the more challenging upsampling by a factor of 4 and the more difficult upsampling by a factor of 8, which correspond with 4x, 16x, and 64x increases in image pixels, respectively. HR images $I_{HR}$ are obtained from the divided dataset, then these images are downsampled by bicubic interpolation to obtain LR images $I_{LR}$, which represent the input for the generator, and all images are normalized to $[0, 1]$. The weighting parameter $\alpha$ for adversarial loss $l_{adv}$ in Equation (\ref{eq:g_loss}) is set to $10^{-3}$ and the weighting parameter $\beta$ for feature loss $l_{fea}$ is set to $10^{-2}$.
	
	The training process can be divided into two parts. We alternate between training the discriminator and the generator, first training the discriminator and then the generator. We first train the discriminator with $l_D$. The initial learning rate of discriminator $lr_D$ is 0.0001. The $lr_D$ is decreased by $1/2$ when the training process is 25\%, 50\%, and 75\% completed. We use Adam \cite{kingma2014adam} with $\beta_1=0.5$, $\beta_2=0.999$ to optimize the discriminator. Whenever a training iteration of the discriminator is completed, a generator is trained with $l_{G}$ immediately afterwards. The initial learning rate of generator $lr_G$ is also set to 0.0001 and decreased by $1/2$ when the training process is completed at 25\%, 50\%, and 75\%. For the generator optimizer, we also use Adam \cite{kingma2014adam} with $\beta_1=0.5$, $\beta_2=0.999$.
	
	For the whole training process, we have carried out more than $10^6$ update iterations both for the generator and the discriminator.

	\begin{figure*}[t]
		\centering
		\includegraphics[width=0.9\linewidth]{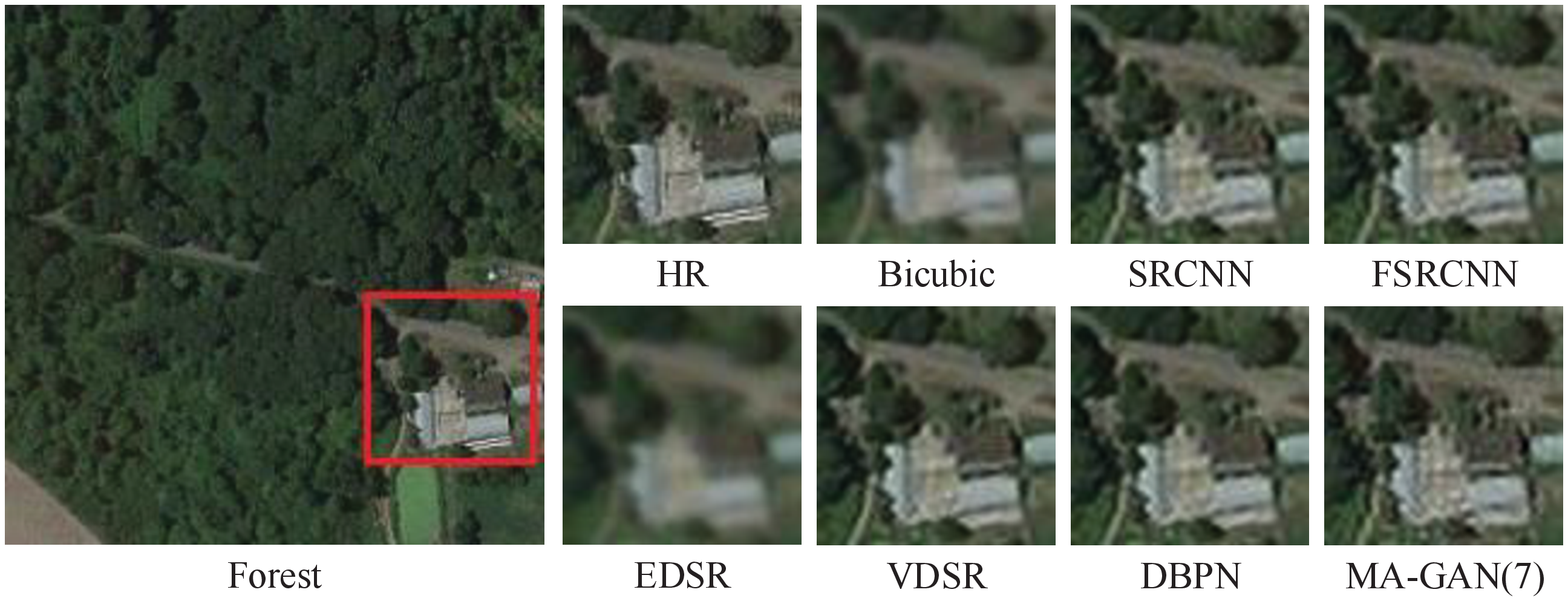}
		\caption{Visual comparisons between MA-GAN and other methods (scale factor $r$=2).}
		\label{fig:result_x2}
	\end{figure*}

	\begin{figure*}[h]
		\centering
		\includegraphics[width=0.9\linewidth]{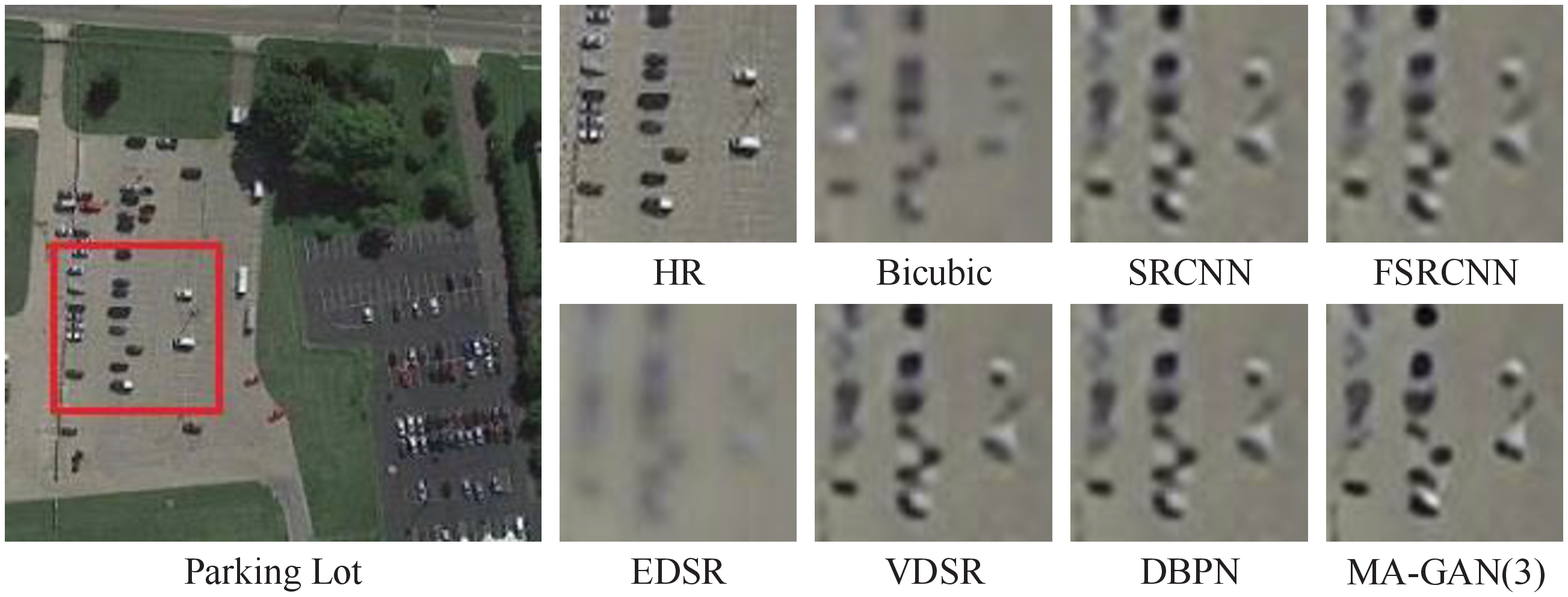}
		\caption{Visual comparisons between MA-GAN and other methods (scale factor $r$=4).}
		\label{fig:result_x4}
	\end{figure*}

	\begin{figure*}[h]
		\centering
		\includegraphics[width=0.9\linewidth]{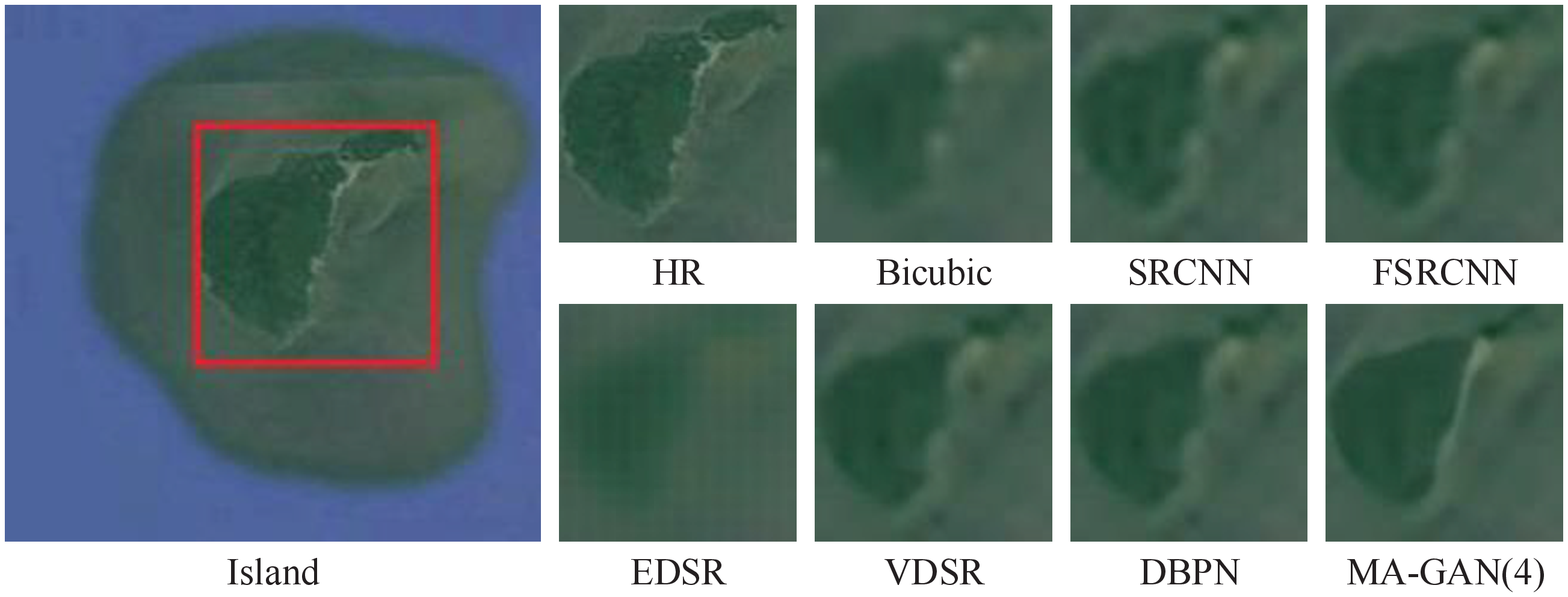}
		\caption{Visual comparisons between MA-GAN and other methods (scale factor $r$=8).}
		\label{fig:result_x8}
	\end{figure*}

	\subsection{Evaluation using different numbers of PCRBD blocks}\label{subsec:n_pcrdb_expe}
	
	\begin{table}[t]
		\centering
		\renewcommand{\arraystretch}{1.1}
		\caption{Quantitative comparison of models using different numbers of PCRBD blocks}	
		\label{tab:n-pcrdb}
		\begin{tabular}{p{0.25\columnwidth}<{\centering} p{0.08\columnwidth}<{\centering} p{0.15\columnwidth}<{\centering} p{0.15\columnwidth}<{\centering}}
			\hline
			\hline
			Model                      & $r$  & PSNR  & SSIM   \\
			\hline
			\multirow{3}{*}{MA-GAN(3)} & 2 & 31.53 & 0.9007 \\
									   & 4 & 27.62 & 0.7640 \\
									   & 8 & 24.50 & 0.5429 \\
			\hline
			\multirow{3}{*}{MA-GAN(4)} & 2 & 31.97 & 0.9101 \\
									   & 4 & 27.76 & 0.7621 \\
									   & 8 & 24.54 & 0.5450 \\
			\hline
			\multirow{3}{*}{MA-GAN(5)} & 2 & 31.37 & 0.8959 \\
									   & 4 & 27.49 & 0.7442 \\
									   & 8 & 24.58 & 0.5409 \\
			\hline
			\multirow{3}{*}{MA-GAN(6)} & 2 & 30.99 & 0.8846 \\
									   & 4 & 27.96 & 0.7535 \\
									   & 8 & 23.85 & 0.5144 \\
			\hline
			\multirow{3}{*}{MA-GAN(7)} & 2 & 31.98 & 0.9102 \\
									   & 4 & 28.05 & 0.7520 \\
									   & 8 & 24.11 & 0.5245 \\
			\hline
			\hline
		\end{tabular}
	\end{table}

	\begin{figure}[h]
		\centering
		\subfigure[The PSNR.]{\label{subfig:psnr}
			\includegraphics[width=0.45\columnwidth]{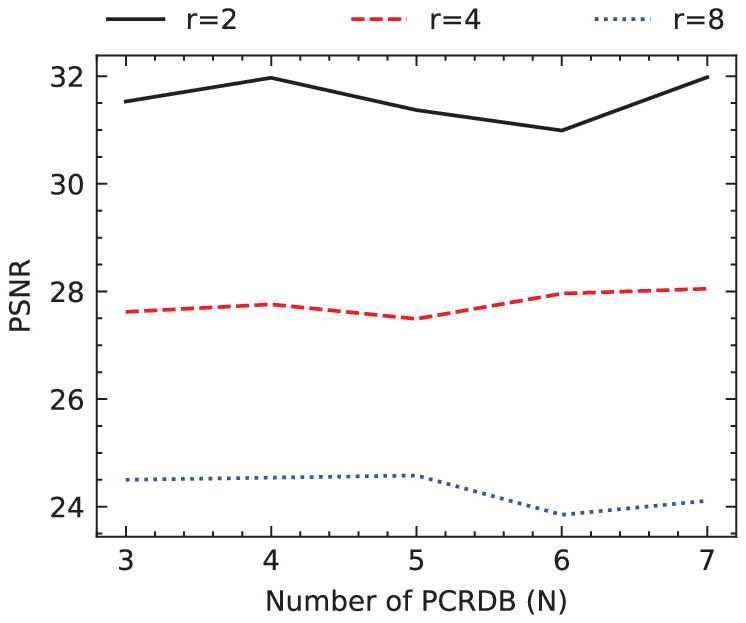}
		}
		\subfigure[The SSIM.]{\label{subfig:ssim}
			\includegraphics[width=0.45\columnwidth]{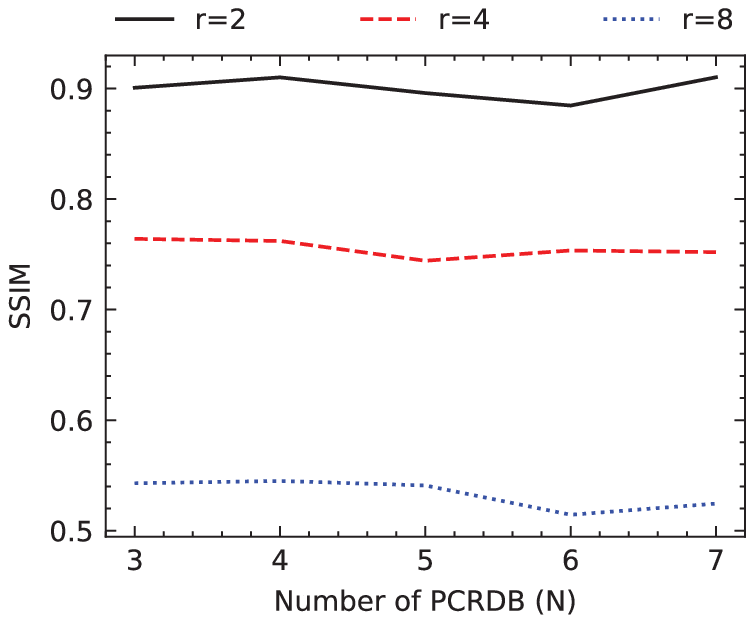}	
		}
		\caption{The comparison results when MA-GAN contains different numbers of PCRDB blocks.}\label{fig:magan}
	\end{figure}
	
	As for the image super-resolution task, the generator that produces the HR images is undoubtedly the most important part. The PCRBD block and AUP block, as the main body of the generator, serve very important roles in it. The number of AUP blocks $S$ is determined by the scale factor $r$ in the experiment. We set each AUP block to upsample by a scale factor of 2 so that $S$ is 1, 2, and 3, corresponding to scale factors of 2, 4, and 8, respectively. Therefore, in this subsection, we conducted experiments on generators using different numbers of PCRDB blocks. We designed experiments with $N\in \left \{ 3,4,5,6,7 \right \}$ for scale factors r of 2, 4, and 8, respectively. We denote the model containing $N$ PCRDB blocks by MA-GAN($N$). To be fair, the PSNR and structural similarity index measure (SSIM) are used as indicators for evaluating quantitative comparison. The quantitative experimental results are shown in Table \ref{tab:n-pcrdb}. We also produced a graph to compare the experimental results, shown in Figure \ref{fig:magan}. 
	
	When $r=2$, MA-GAN(7) reaches the highest PSNR value of 31.98 dB. The lowest value of 30.99 dB corresponds with MA-GAN(4), a difference of 0.99 dB. Additionally, the SSIM values obtained for these two models are 0.9102 and 0.8846, respectively, with a difference of 0.0256. In regard to $r=4$, MA-GAN(7) reaches the highest PSNR value of 28.05 dB. The lowest is MA-GAN(5) at 27.49 dB, a difference of 0.56 dB. The highest SSIM at this point is that of MA-GAN(3) at 0.7640, while the lowest is that of MA-GAN(5) at 0.7442, a difference of 0.0198. Finally, for $r=8$, MA-GAN(5) obtained the highest PSNR value of 24.58 dB, while the lowest was that of MA-GAN(6) at 23.85 dB, a difference of 0.73 dB. For SSIM, MA-GAN(4) obtained the highest value of 0.5450, and the lowest was that of MA-GAN(6) at 0.5144, a difference of 0.0306. It can also be more intuitively determined from Figure \ref{fig:magan} that the performances of the different models do not differ much when $N$ varies. This is due to our multi-attention mechanism, which allows the model to automatically adjust the input of the corresponding feature maps to achieve better SR performance.

	\subsection{Comparison with different methods} \label{subsec:exp_results}

	\begin{table*}[]
		\caption{Quantitative comparison of the different SR methods on the dataset (scale factor $r$=2).}
		\centering
		\centering\renewcommand\arraystretch{1.1}
		\label{tab:results_x2}
		\begin{tabular}{p{0.08\columnwidth}<{\centering}|p{0.075\columnwidth}p{0.1\columnwidth}<{\centering}|p{0.075\columnwidth}p{0.1\columnwidth}<{\centering}|p{0.075\columnwidth}p{0.1\columnwidth}<{\centering}|p{0.075\columnwidth}p{0.1\columnwidth}<{\centering}|p{0.075\columnwidth}p{0.1\columnwidth}<{\centering}|p{0.075\columnwidth}p{0.1\columnwidth}<{\centering}|p{0.075\columnwidth}p{0.1\columnwidth}<{\centering}}
			\hline
			\hline
			\multirow{2}{*}{Scene} &
			\multicolumn{2}{c|}{Bicubic} &
			\multicolumn{2}{c|}{SRCNN} &
			\multicolumn{2}{c|}{FSRCNN} &
			\multicolumn{2}{c|}{VDSR} &
			\multicolumn{2}{c|}{EDSR} &
			\multicolumn{2}{c|}{DBPN} &
			\multicolumn{2}{c}{MA-GAN(7)} \\
			\cline{2-15}
			& PSNR    & SSIM   & PSNR  & SSIM   & PSNR   & SSIM   & PSNR  & SSIM   & PSNR  & SSIM   & PSNR  & SSIM   & PSNR      & SSIM   \\
			\hline
			S1      & 25.27   & 0.8338 & 25.07 & 0.8400 & 24.69  & 0.8354 & 25.20 & \textbf{0.8657} & 21.57 & 0.6381 & 25.38 & 0.8635 & \textbf{27.61}     & 0.8613 \\
			S2      & 33.36   & 0.9207 & 33.06 & 0.9194 & 32.17  & 0.9182 & 33.65 & \textbf{0.9385} & 27.06 & 0.8300 & 33.60 & 0.9377 & \textbf{34.11}     & 0.9253 \\
			S3      & \textbf{38.57}   & 0.9585 & 36.33 & 0.9578 & 35.44  & 0.9570 & 36.26 & 0.9610 & 31.09 & 0.9037 & 36.34 & \textbf{0.9642} & 37.13     & 0.9462 \\
			S4      & 30.18   & 0.8477 & 29.69 & 0.8667 & 29.45  & 0.8657 & 29.82 & 0.8755 & 26.25 & 0.7064 & 29.92 & 0.8779 & \textbf{32.26}    & \textbf{0.8865} \\
			S5      & 27.25   & 0.8714 & 27.98 & 0.8897 & 27.74  & 0.8879 & 28.22 & 0.9062 & 24.34 & 0.7371 & 28.31 & \textbf{0.9073} & \textbf{29.29}     & 0.8971 \\
			S6      & 32.82   & 0.9071 & 31.37 & 0.9019 & 30.91  & 0.9011 & 31.54 & 0.9125 & 26.70 & 0.7537 & 31.61 & 0.9140 & \textbf{34.00}     & \textbf{0.9243} \\
			S7      & 28.69   & 0.8598 & 28.47 & 0.8745 & 28.29  & 0.8735 & 28.84 & \textbf{0.8929} & 25.05 & 0.7278 & 28.87 & 0.8924 & \textbf{30.95}     & 0.8926 \\
			S8      & 29.57   & 0.9006 & 29.21 & 0.9060 & 28.74  & 0.9042 & 29.39 & 0.9186 & 24.25 & 0.7461 & 29.48 & \textbf{0.9191} & \textbf{30.93}     & 0.9158 \\
			S9      & 33.61   & 0.9043 & 31.97 & 0.9002 & 31.74  & 0.9000 & 31.86 & 0.9018 & 28.45 & 0.7576 & 32.00 & 0.9063 & \textbf{34.78}     & \textbf{0.9227} \\
			S10     & \textbf{35.41}   & 0.9184 & 34.11 & 0.9260 & 32.55  & 0.9246 & 33.95 & 0.9314 & 25.12 & 0.7893 & 34.07 & \textbf{0.9332} & 34.05     & 0.9214 \\
			S11     & 29.72   & 0.9001 & 29.01 & 0.9086 & 28.62  & 0.9068 & 29.21 & 0.9219 & 24.45 & 0.7605 & 29.32 & \textbf{0.9229} & \textbf{31.26}     & 0.9174 \\
			S12     & 29.21   & 0.8618 & 28.88 & 0.8733 & 28.62  & 0.8719 & 29.33 & \textbf{0.8936} & 25.36 & 0.7181 & 29.33 & 0.8922 & \textbf{31.31}     & 0.8928 \\
			S13     & 26.63   & 0.8928 & 26.70 & 0.9068 & 26.23  & 0.9026 & 26.64 & 0.9154 & 21.76 & 0.7378 & 26.77 & \textbf{0.9164} & \textbf{28.46}     & 0.9110 \\
			S14     & 28.18   & 0.8634 & 28.16 & 0.8761 & 27.98  & 0.8749 & 28.42 & 0.8910 & 24.67 & 0.7155 & 28.52 & 0.8918 & \textbf{30.34}     & \textbf{0.8953} \\
			S15     & 30.62   & 0.8680 & 29.64 & 0.8671 & 29.42  & 0.8664 & 29.89 & 0.8812 & 26.25 & 0.7121 & 29.98 & 0.8817 & \textbf{32.48}     & \textbf{0.8995} \\
			S16     & 29.45   & 0.8997 & 29.29 & 0.9069 & 29.03  & 0.9062 & 29.61 & 0.9189 & 25.17 & 0.7615 & 29.66 & \textbf{0.9202} & \textbf{31.30}     & 0.9180 \\
			S17     & 28.24   & 0.8780 & 27.24 & 0.8660 & 26.96  & 0.8640 & 27.46 & 0.8863 & 23.70 & 0.7026 & 27.57 & 0.8860 & \textbf{29.92}     & \textbf{0.8969} \\
			S18     & 26.15   & 0.8717 & 26.25 & 0.8826 & 25.95  & 0.8800 & 26.62 & \textbf{0.9014} & 22.19 & 0.7097 & 26.72 & 0.9013 & \textbf{28.20}     & 0.8931 \\
			S19     & 33.66   & 0.9176 & 32.53 & 0.9177 & 32.12  & 0.9168 & 32.50 & 0.9213 & 27.98 & 0.7974 & 32.65 & 0.9253 & \textbf{34.49}     & \textbf{0.9287} \\
			S20     & 33.03   & 0.8977 & 32.33 & 0.9011 & 31.81  & 0.9006 & 32.65 & 0.9153 & 27.50 & 0.7923 & 32.74 & \textbf{0.9175} & \textbf{34.34}     & 0.9151 \\
			S21     & 31.79   & 0.8997 & 31.11 & 0.9077 & 30.70  & 0.9072 & 31.04 & 0.9100 & 26.43 & 0.7461 & 31.18 & 0.9137 & \textbf{33.18}     & \textbf{0.9216} \\
			S22     & 32.30   & 0.9214 & 31.31 & 0.9192 & 30.99  & 0.9187 & 31.49 & 0.9303 & 27.35 & 0.8276 & 31.56 & \textbf{0.9322} & \textbf{33.86}     & 0.9295 \\
			S23     & 33.47   & 0.9156 & 31.92 & 0.9117 & 31.49  & 0.9113 & 31.94 & 0.9172 & 27.40 & 0.7774 & 32.08 & 0.9208 & \textbf{34.42}     & \textbf{0.9273} \\
			S24     & 31.09   & 0.9058 & 30.60 & 0.9073 & 30.30  & 0.9065 & 30.92 & 0.9205 & 26.75 & 0.7924 & 30.99 & \textbf{0.9226} & \textbf{32.75}     & 0.9213 \\
			S25     & 27.27   & 0.8632 & 27.27 & 0.8745 & 26.95  & 0.8721 & 27.44 & \textbf{0.8968} & 23.33 & 0.7019 & 27.66 & 0.8966 & \textbf{29.25}     & 0.8857 \\
			S26     & 34.62   & 0.8763 & 33.52 & 0.8831 & 33.05  & 0.8832 & 33.36 & 0.8817 & 29.85 & 0.7406 & 33.55 & 0.8885 & \textbf{35.74}     & \textbf{0.9063} \\
			S27     & 33.65   & 0.9274 & 33.04 & 0.9217 & 32.36  & 0.9207 & 33.55 & 0.9380 & 27.27 & 0.8287 & 33.58 & \textbf{0.9393} & \textbf{34.73}     & 0.9325 \\
			S28     & 29.99   & 0.8711 & 29.20 & 0.8689 & 28.94  & 0.8679 & 29.52 & 0.8893 & 25.70 & 0.7139 & 29.60 & 0.8891 & \textbf{31.73}     & \textbf{0.8939} \\
			S29     & 32.02   & 0.9000 & 30.79 & 0.9022 & 30.23  & 0.9010 & 30.87 & 0.9101 & 26.02 & 0.7801 & 30.94 & 0.9122 & \textbf{33.38}     & \textbf{0.9177} \\
			S30     & 33.61   & 0.9430 & 31.72 & 0.9270 & 31.22  & 0.9231 & 31.71 & 0.9362 & 26.88 & 0.8162 & 31.91 & \textbf{0.9403} & \textbf{33.97}     & 0.9354 \\
			S31     & 27.66   & 0.8581 & 27.35 & 0.8700 & 27.14  & 0.8686 & 27.65 & 0.8889 & 23.93 & 0.7047 & 27.70 & 0.8879 & \textbf{29.87}     & \textbf{0.8901} \\
			S32     & 26.09   & 0.8608 & 25.98 & 0.8731 & 25.81  & 0.8712 & 26.26 & \textbf{0.8930} & 22.53 & 0.6897 & 26.34 & 0.8923 & \textbf{28.33}     & 0.8913 \\
			S33     & 27.71   & 0.8564 & 27.80 & 0.8690 & 27.61  & 0.8668 & 27.66 & 0.8773 & 23.78 & 0.6890 & 27.85 & 0.8791 & \textbf{29.70}     & \textbf{0.8853} \\
			S34     & 29.22   & 0.8809 & 28.11 & 0.8873 & 27.82  & 0.8862 & 28.26 & 0.8979 & 24.06 & 0.6997 & 28.34 & 0.8984 & \textbf{31.00}     & \textbf{0.9074} \\
			S35     & 30.26   & 0.9132 & 30.20 & 0.9145 & 29.71  & 0.9131 & 30.34 & 0.9236 & 25.15 & 0.7632 & 30.43 & 0.9252 & \textbf{31.53}     & \textbf{0.9260} \\
			S36     & 25.32   & 0.8822 & 26.16 & 0.8902 & 25.77  & 0.8855 & 26.24 & 0.9117 & 22.35 & 0.7578 & 26.45 & \textbf{0.9123} & \textbf{27.63}     & 0.8897 \\
			S37     & 29.62   & 0.9064 & 28.13 & 0.8924 & 27.79  & 0.8902 & 28.45 & 0.9102 & 23.84 & 0.7391 & 28.50 & 0.9098 & \textbf{30.99}     & \textbf{0.9193} \\
			S38     & 30.03   & 0.8474 & 29.88 & 0.8645 & 29.86  & 0.8646 & 29.55 & 0.8593 & 27.31 & 0.6602 & 29.75 & 0.8650 & \textbf{32.00}     & \textbf{0.8866} \\
			S39     & 29.06   & 0.8546 & 28.66 & 0.8622 & 28.42  & 0.8615 & 28.88 & 0.8764 & 25.10 & 0.6852 & 28.94 & 0.8761 & \textbf{31.09}     & \textbf{0.8935} \\
			S40     & 30.61   & 0.9041 & 29.84 & 0.9019 & 29.43  & 0.9007 & 29.98 & 0.9122 & 25.33 & 0.7470 & 30.08 & 0.9139 & \textbf{32.15}     & \textbf{0.9199} \\
			S41     & 27.69   & 0.8793 & 27.81 & 0.8867 & 27.54  & 0.8849 & 28.06 & 0.9028 & 23.92 & 0.7272 & 28.14 & \textbf{0.9032} & \textbf{29.61}     & 0.9007 \\
			S42     & 31.06   & 0.9148 & 31.56 & 0.9189 & 30.97  & 0.9176 & 31.67 & 0.9283 & 26.67 & 0.8126 & 31.75 & \textbf{0.9301} & \textbf{32.56}     & 0.9213 \\
			S43     & 32.92   & 0.9205 & 32.17 & 0.9187 & 31.80  & 0.9179 & 32.38 & 0.9275 & 28.08 & 0.8181 & 32.51 & \textbf{0.9309} & \textbf{34.29}     & 0.9303 \\
			S44     & 31.09   & 0.9108 & 30.99 & 0.9095 & 30.23  & 0.9075 & 31.32 & \textbf{0.9281} & 24.93 & 0.7947 & 31.41 & 0.9272 & \textbf{32.03}     & 0.9137 \\
			S45     & \textbf{37.08}   & 0.9535 & 36.21 & 0.9537 & 34.69  & 0.9527 & 35.97 & 0.9557 & 28.30 & 0.8773 & 36.09 & \textbf{0.9588} & 36.13     & 0.9517 \\
			\hline
			Avg     & 30.55   & 0.8920 & 29.97 & 0.8961 & 29.54  & 0.8946 & 30.12 & 0.9083 & 25.58 & 0.7531 & 30.23 & 0.9097 &\textbf{31.98} &\textbf{0.9102} \\
			\hline
			\hline
		\end{tabular}
	\end{table*}

	\begin{table*}[]
		\caption{Quantitative comparison of the different SR methods on the dataset (scale factor $r$=4).}
		\centering
		\centering\renewcommand\arraystretch{1.1}
		\label{tab:results_x4}
		\begin{tabular}{p{0.08\columnwidth}<{\centering}|p{0.075\columnwidth}p{0.1\columnwidth}<{\centering}|p{0.075\columnwidth}p{0.1\columnwidth}<{\centering}|p{0.075\columnwidth}p{0.1\columnwidth}<{\centering}|p{0.075\columnwidth}p{0.1\columnwidth}<{\centering}|p{0.075\columnwidth}p{0.1\columnwidth}<{\centering}|p{0.075\columnwidth}p{0.1\columnwidth}<{\centering}|p{0.075\columnwidth}p{0.1\columnwidth}<{\centering}}
			\hline
			\hline
			\multirow{2}{*}{Scene} &
			\multicolumn{2}{c|}{Bicubic} &
			\multicolumn{2}{c|}{SRCNN} &
			\multicolumn{2}{c|}{FSRCNN} &
			\multicolumn{2}{c|}{VDSR} &
			\multicolumn{2}{c|}{EDSR} &
			\multicolumn{2}{c|}{DBPN} &
			\multicolumn{2}{c}{MA-GAN(3)} \\
			\cline{2-15}
			& PSNR    & SSIM   & PSNR  & SSIM   & PSNR   & SSIM   & PSNR  & SSIM   & PSNR  & SSIM   & PSNR  & SSIM   & PSNR      & SSIM   \\
			\hline
			S1  & 19.72 & 0.5335 & 21.33 & 0.6073 & 21.24 & 0.6056 & 21.39 & 0.6222 & 19.13 & 0.4013 & 21.28 & 0.6379 & \textbf{23.11} & \textbf{0.6506} \\
			S2  & 27.21 & 0.7520 & 29.90 & 0.8178 & 29.64 & 0.8172 & 30.08 & 0.8314 & 24.30 & 0.6778 & 30.34 & 0.8493 & \textbf{30.64} & \textbf{0.8553} \\
			S3  & 32.03 & 0.8663 & 33.47 & 0.9003 & 33.10 & 0.9008 & 33.23 & 0.9025 & 28.30 & 0.8019 & 32.95 & 0.9079 & \textbf{35.40} & \textbf{0.9079} \\
			S4  & 24.88 & 0.5967 & 26.63 & 0.6832 & 26.53 & 0.6814 & 26.69 & 0.6883 & 23.66 & 0.5094 & \textbf{26.71} & 0.6993 & 26.54 & \textbf{0.7246} \\
			S5  & 21.36 & 0.6011 & 24.24 & 0.7236 & 24.10 & 0.7204 & 24.23 & 0.7308 & 21.10 & 0.4874 & 24.21 & \textbf{0.7449} & \textbf{26.81} & 0.7446 \\
			S6  & 26.33 & 0.6688 & 27.85 & 0.7405 & 27.69 & 0.7381 & 27.81 & 0.7462 & 23.92 & 0.5293 & 27.95 & 0.7639 & \textbf{29.16} & \textbf{0.7956} \\
			S7  & 23.37 & 0.6115 & 25.18 & 0.7119 & 25.06 & 0.7088 & 25.24 & 0.7208 & 22.12 & 0.5044 & 25.27 & 0.7361 & \textbf{26.32} & \textbf{0.7422} \\
			S8  & 22.96 & 0.6397 & 25.08 & 0.7372 & 24.92 & 0.7337 & 25.06 & 0.7428 & 20.93 & 0.4810 & 25.11 & \textbf{0.7584} & \textbf{25.42} & 0.7093 \\
			S9  & 27.61 & 0.6779 & 28.58 & 0.7372 & 28.51 & 0.7373 & 28.56 & 0.7403 & 25.79 & 0.5548 & 28.51 & 0.7467 & \textbf{29.83} & \textbf{0.7520} \\
			S10 & 29.02 & 0.7170 & 30.70 & 0.7891 & 30.33 & 0.7903 & 30.46 & 0.7905 & 22.40 & 0.5884 & 30.25 & 0.7959 & \textbf{31.69} & \textbf{0.8448} \\
			S11 & 23.31 & 0.6544 & 25.05 & 0.7521 & 24.87 & 0.7482 & 24.98 & 0.7577 & 20.99 & 0.5032 & 24.93 & \textbf{0.7709} & \textbf{25.83} & 0.7225 \\
			S12 & 23.59 & 0.6076 & 25.71 & 0.7057 & 25.57 & 0.7019 & 25.78 & 0.7143 & 22.50 & 0.4913 & 26.07 & 0.7383 & \textbf{26.56} & \textbf{0.7457} \\
			S13 & 20.45 & 0.6206 & 22.56 & 0.7350 & 22.40 & 0.7303 & 22.47 & 0.7369 & 18.31 & 0.4580 & 22.06 & \textbf{0.7376} & \textbf{23.13} & 0.6996 \\
			S14 & 22.57 & 0.6005 & 24.63 & 0.6977 & 24.53 & 0.6952 & 24.67 & 0.7053 & 21.69 & 0.4723 & 24.67 & \textbf{0.7191} & \textbf{25.45} & 0.6943 \\
			S15 & 24.99 & 0.6154 & 26.68 & 0.6984 & 26.58 & 0.6961 & 26.72 & 0.7050 & 23.59 & 0.5035 & 26.90 & 0.7238 & \textbf{27.77} & \textbf{0.7336} \\
			S16 & 23.36 & 0.6541 & 25.54 & 0.7512 & 25.41 & 0.7490 & 25.54 & 0.7566 & 22.02 & 0.5156 & 25.56 & 0.7692 & \textbf{27.94} & \textbf{0.7776} \\
			S17 & 21.92 & 0.6152 & 23.78 & 0.6873 & 23.65 & 0.6837 & 23.84 & 0.6985 & 20.74 & 0.4642 & 23.80 & 0.7150 & \textbf{26.53} & \textbf{0.7754} \\
			S18 & 19.93 & 0.5891 & 22.27 & 0.6984 & 22.10 & 0.6941 & 22.30 & 0.7082 & 18.83 & 0.4374 & 22.21 & 0.7218 & \textbf{23.68} & \textbf{0.7460} \\
			S19 & 27.45 & 0.7115 & 29.13 & 0.7854 & 28.99 & 0.7847 & 29.04 & 0.7867 & 25.19 & 0.6058 & 28.90 & 0.7906 & \textbf{30.88} & \textbf{0.8149} \\
			S20 & 27.09 & 0.7197 & 29.42 & 0.7850 & 29.22 & 0.7826 & 29.45 & 0.7945 & 24.93 & 0.6237 & \textbf{29.60} & \textbf{0.8130} & 29.33 & 0.7694 \\
			S21 & 25.59 & 0.6384 & 27.45 & 0.7378 & 27.33 & 0.7371 & 27.31 & 0.7355 & 23.65 & 0.4988 & 27.29 & 0.7411 & \textbf{28.54} & \textbf{0.7476} \\
			S22 & 26.65 & 0.7682 & 28.11 & 0.8196 & 27.92 & 0.8170 & 28.06 & 0.8259 & 24.47 & 0.6725 & 28.06 & 0.8366 & \textbf{30.75} & \textbf{0.8459} \\
			S23 & 27.15 & 0.7002 & 28.44 & 0.7653 & 28.32 & 0.7646 & 28.34 & 0.7664 & 24.58 & 0.5658 & 28.34 & 0.7759 & \textbf{29.12} & \textbf{0.7904} \\
			S24 & 25.19 & 0.7013 & 27.21 & 0.7790 & 27.08 & 0.7767 & 27.27 & 0.7866 & 23.75 & 0.6000 & 27.35 & 0.8008 & \textbf{29.35} & \textbf{0.8035} \\
			S25 & 21.12 & 0.5777 & 23.39 & 0.6800 & 23.26 & 0.6757 & 23.44 & 0.6918 & 20.49 & 0.4590 & 23.47 & 0.7124 & \textbf{25.98} & \textbf{0.7243} \\
			S26 & 29.14 & 0.6360 & 30.98 & 0.7089 & 30.84 & 0.7087 & 30.84 & 0.7076 & 28.12 & 0.5960 & 30.78 & 0.7130 & \textbf{31.38} & \textbf{0.7352} \\
			S27 & 27.56 & 0.7520 & 30.08 & 0.8223 & 29.85 & 0.8208 & 30.16 & 0.8325 & 24.58 & 0.6605 & \textbf{30.21} & \textbf{0.8469} & 29.61 & 0.7876 \\
			S28 & 24.13 & 0.6170 & 26.06 & 0.6871 & 25.95 & 0.6845 & 26.16 & 0.6974 & 23.19 & 0.5191 & 26.24 & 0.7158 & \textbf{27.58} & \textbf{0.7342} \\
			S29 & 26.29 & 0.7179 & 27.74 & 0.7696 & 27.56 & 0.7690 & 27.67 & 0.7741 & 23.13 & 0.6071 & 27.55 & 0.7806 & \textbf{29.02} & \textbf{0.8066} \\
			S30 & 25.99 & 0.7586 & 27.32 & 0.8002 & 27.06 & 0.7941 & 27.16 & 0.8052 & 22.83 & 0.5885 & 27.05 & 0.8139 & \textbf{29.02} & \textbf{0.8304} \\
			S31 & 22.15 & 0.6004 & 24.01 & 0.6939 & 23.91 & 0.6913 & 24.04 & 0.7029 & 21.07 & 0.4572 & 24.13 & 0.7211 & \textbf{25.23} & \textbf{0.7602} \\
			S32 & 20.30 & 0.5723 & 22.28 & 0.6812 & 22.14 & 0.6770 & 22.27 & 0.6887 & 19.40 & 0.4036 & 22.17 & 0.7023 & \textbf{23.22} & \textbf{0.7146} \\
			S33 & 21.77 & 0.5602 & 24.15 & 0.6680 & 24.10 & 0.6685 & \textbf{24.26} & 0.6806 & 20.73 & 0.4311 & 24.25 & 0.6953 & 23.72 & \textbf{0.7145} \\
			S34 & 23.13 & 0.6006 & 24.39 & 0.6922 & 24.24 & 0.6872 & 24.31 & 0.6940 & 21.00 & 0.4201 & 24.37 & 0.7111 & \textbf{27.96} & \textbf{0.7541} \\
			S35 & 23.45 & 0.6689 & \textbf{26.12} & 0.7523 & 25.95 & 0.7496 & 26.03 & 0.7554 & 21.85 & 0.5086 & 26.03 & \textbf{0.7673} & 25.85 & 0.7336 \\
			S36 & 19.50 & 0.6524 & 22.27 & 0.7446 & 22.09 & 0.7411 & 22.24 & 0.7601 & 19.41 & 0.5517 & 22.09 & \textbf{0.7772} & \textbf{22.54} & 0.6968 \\
			S37 & 22.87 & 0.6695 & 24.49 & 0.7321 & 24.34 & 0.7270 & 24.53 & 0.7401 & 20.44 & 0.4843 & 24.47 & \textbf{0.7565} & \textbf{26.90} & 0.7561 \\
			S38 & 24.57 & 0.5283 & 26.73 & 0.6347 & 26.71 & 0.6330 & 26.68 & 0.6345 & 24.99 & 0.4232 & 26.63 & 0.6403 & \textbf{27.54} & \textbf{0.6499} \\
			S39 & 23.39 & 0.5813 & 25.54 & 0.6748 & 25.40 & 0.6703 & 25.52 & 0.6806 & 22.54 & 0.4518 & 25.68 & 0.7049 & \textbf{26.20} & \textbf{0.7405} \\
			S40 & 24.36 & 0.6606 & 26.23 & 0.7377 & 26.11 & 0.7360 & 26.25 & 0.7447 & 22.32 & 0.5097 & 26.29 & 0.7587 & \textbf{27.99} & \textbf{0.7965} \\
			S41 & 21.59 & 0.6107 & 24.08 & 0.7162 & 23.94 & 0.7125 & 24.07 & 0.7235 & 20.65 & 0.4670 & 24.02 & 0.7371 & \textbf{27.33} & \textbf{0.7443} \\
			S42 & 25.13 & 0.7546 & \textbf{28.15} & 0.8047 & 27.93 & 0.8033 & 28.06 & 0.8097 & 23.80 & 0.6489 & 27.94 & 0.8192 & 26.29 & \textbf{0.8265} \\
			S43 & 26.83 & 0.7304 & 28.85 & 0.8043 & 28.73 & 0.8036 & 28.94 & 0.8121 & 24.93 & 0.6441 & 28.92 & \textbf{0.8216} & \textbf{29.07} & 0.7798 \\
			S44 & 24.39 & 0.7083 & 27.28 & 0.7876 & 27.01 & 0.7859 & 27.34 & 0.7993 & 21.92 & 0.6074 & 27.34 & 0.8155 & \textbf{29.10} & \textbf{0.8211} \\
			S45 & 30.07 & 0.8142 & \textbf{32.36} & 0.8695 & 31.98 & 0.8714 & 32.01 & 0.8692 & 24.76 & 0.7233 & 31.84 & 0.8753 & 31.67 & \textbf{0.8806} \\
			\hline
			Avg & 24.48 & 0.6585 & 26.48 & 0.7402 & 26.31 & 0.7379 & 26.46 & 0.7466 & 22.65 & 0.5358 & 26.44 & 0.7596 & \textbf{27.62} & \textbf{0.7640} \\	
			\hline
			\hline
		\end{tabular}
	\end{table*}

	\begin{table*}[]
		\caption{Quantitative comparison of the different SR methods on the dataset (scale factor $r$=8).}
		\centering
		\centering\renewcommand\arraystretch{1.1}
		\label{tab:results_x8}
		\begin{tabular}{p{0.08\columnwidth}<{\centering}|p{0.075\columnwidth}p{0.1\columnwidth}<{\centering}|p{0.075\columnwidth}p{0.1\columnwidth}<{\centering}|p{0.075\columnwidth}p{0.1\columnwidth}<{\centering}|p{0.075\columnwidth}p{0.1\columnwidth}<{\centering}|p{0.075\columnwidth}p{0.1\columnwidth}<{\centering}|p{0.075\columnwidth}p{0.1\columnwidth}<{\centering}|p{0.075\columnwidth}p{0.1\columnwidth}<{\centering}}
			\hline
			\hline
			\multirow{2}{*}{Scene} &
			\multicolumn{2}{c|}{Bicubic} &
			\multicolumn{2}{c|}{SRCNN} &
			\multicolumn{2}{c|}{FSRCNN} &
			\multicolumn{2}{c|}{VDSR} &
			\multicolumn{2}{c|}{EDSR} &
			\multicolumn{2}{c|}{DBPN} &
			\multicolumn{2}{c}{MA-GAN(4)} \\
			\cline{2-15}
				  & PSNR    & SSIM   & PSNR  & SSIM   & PSNR   & SSIM   & PSNR  & SSIM   & PSNR  & SSIM   & PSNR  & SSIM   & PSNR      & SSIM   \\
			\hline
			S1  & 17.52 & 0.3522 & 19.27 & 0.3731 & 19.06 & 0.3795 & 19.34 & 0.3873 & 17.71 & 0.2986 & 19.19 & 0.4029 & \textbf{20.69} & \textbf{0.4250} \\
			S2  & 24.37 & 0.6217 & 26.69 & 0.6639 & 25.49 & 0.6692 & 26.87 & 0.6834 & 21.99 & 0.5903 & 27.03 & 0.7013 & \textbf{29.62} & \textbf{0.7586} \\
			S3  & 28.33 & 0.7773 & 30.04 & 0.7926 & 28.48 & 0.7895 & 30.11 & 0.8035 & 26.22 & 0.7316 & \textbf{30.13} & \textbf{0.8137} & 29.88 & 0.7787 \\
			S4  & 22.60 & 0.4542 & 24.24 & 0.4965 & 23.73 & 0.4951 & 24.25 & 0.5023 & 21.79 & 0.4059 & 24.15 & 0.5095 & \textbf{25.50} & \textbf{0.5325} \\
			S5  & 18.43 & 0.3721 & 21.28 & 0.4649 & 20.85 & 0.4734 & \textbf{21.31} & 0.4808 & 18.85 & 0.3303 & 20.96 & \textbf{0.4870} & 21.30 & 0.4412 \\
			S6  & 23.24 & 0.4714 & 24.96 & 0.5132 & 24.48 & 0.5154 & 25.01 & 0.5204 & 22.00 & 0.4031 & 24.80 & 0.5256 & \textbf{25.86} & \textbf{0.5335} \\
			S7  & 20.71 & 0.4255 & 22.36 & 0.4827 & 22.01 & 0.4853 & 22.41 & 0.4927 & 20.01 & 0.3671 & 22.18 & \textbf{0.4993} & \textbf{23.57} & 0.4966 \\
			S8  & 19.58 & 0.3966 & 21.63 & 0.4597 & 21.14 & 0.4638 & 21.65 & 0.4654 & 18.60 & 0.3103 & 21.26 & \textbf{0.4688} & \textbf{22.17} & 0.4547 \\
			S9  & 24.74 & 0.4978 & 26.19 & 0.5392 & 25.24 & 0.5334 & 26.15 & 0.5422 & 24.22 & 0.4411 & 25.94 & 0.5433 & \textbf{27.21} & \textbf{0.5573} \\
			S10 & 26.06 & 0.5682 & 27.94 & 0.6119 & 23.60 & 0.6014 & 27.94 & 0.6162 & 20.12 & 0.5014 & 27.79 & 0.6139 & \textbf{28.39} & \textbf{0.6271} \\
			S11 & 19.89 & 0.4222 & 21.54 & 0.4794 & 21.29 & 0.4895 & 21.62 & 0.4944 & 18.46 & 0.3363 & 21.30 & 0.5022 & \textbf{24.39} & \textbf{0.5220} \\
			S12 & 20.84 & 0.4154 & 22.91 & 0.4666 & 22.63 & 0.4743 & 23.04 & 0.4826 & 20.38 & 0.3574 & 23.05 & \textbf{0.5034} & \textbf{23.71} & 0.4869 \\
			S13 & 17.25 & 0.3719 & 19.43 & 0.4616 & 18.49 & 0.4613 & 19.32 & 0.4625 & 15.71 & 0.2736 & 18.78 & 0.4618 & \textbf{20.22} & \textbf{0.5358} \\
			S14 & 19.79 & 0.3907 & 21.88 & 0.4554 & 21.57 & 0.4557 & 21.85 & 0.4598 & 19.51 & 0.3199 & 21.49 & 0.4573 & \textbf{22.54} & \textbf{0.4619}\\
			S15 & 22.38 & 0.4420 & 24.18 & 0.4887 & 23.74 & 0.4915 & 24.25 & 0.4984 & 21.74 & 0.3885 & 24.18 & 0.5100 & \textbf{25.17} & \textbf{0.5169} \\
			S16 & 20.21 & 0.4267 & 22.35 & 0.4908 & 22.05 & 0.4961 & 22.40 & 0.4989 & 19.87 & 0.3625 & 22.10 & \textbf{0.5075} & \textbf{22.89} & 0.4818 \\
			S17 & 18.87 & 0.4017 & 21.07 & 0.4444 & 20.71 & 0.4561 & 21.14 & 0.4645 & 18.65 & 0.3292 & 20.85 & 0.4715 & \textbf{23.07} & \textbf{0.5120} \\
			S18 & 16.84 & 0.3613 & 18.95 & 0.4157 & 18.79 & 0.4230 & 18.99 & 0.4249 & 16.64 & 0.2714 & 18.71 & 0.4334 & \textbf{20.53} & \textbf{0.4894} \\
			S19 & 24.44 & 0.5449 & 26.18 & 0.5895 & 25.05 & 0.5913 & 26.24 & 0.6007 & 22.91 & 0.4870 & 26.02 & 0.6036 & \textbf{27.78} & \textbf{0.6290} \\
			S20 & 24.08 & 0.5774 & 26.41 & 0.6018 & 25.86 & 0.6118 & 26.56 & 0.6205 & 22.91 & 0.5286 & 26.72 & 0.6481 & \textbf{28.53} & \textbf{0.6608} \\
			S21 & 22.58 & 0.4159 & 24.67 & \textbf{0.4947} & 24.09 & 0.4917 & 24.63 & 0.4927 & 21.68 & 0.3555 & 24.23 & 0.4829 & \textbf{25.21} & 0.4877 \\
			S22 & 23.78 & 0.6420 & 25.09 & 0.6484 & 24.35 & 0.6553 & 25.28 & 0.6713 & 22.24 & 0.5742 & 25.21 & 0.6919 & \textbf{27.63} & \textbf{0.7302} \\
			S23 & 24.10 & 0.5203 & 25.51 & 0.5472 & 24.78 & 0.5499 & 25.54 & 0.5538 & 22.53 & 0.4377 & 25.27 & 0.5546 & \textbf{26.57} & \textbf{0.5764} \\
			S24 & 22.20 & 0.5272 & 24.20 & 0.5738 & 23.69 & 0.5818 & 24.31 & 0.5899 & 21.69 & 0.4833 & 24.20 & 0.6047 & \textbf{26.46} & \textbf{0.6281} \\
			S25 & 18.57 & 0.3737 & 20.71 & 0.4291 & 20.51 & 0.4382 & 20.80 & 0.4465 & 18.56 & 0.3293 & 20.62 & 0.4589 & \textbf{22.22} & \textbf{0.4739} \\
			S26 & 27.24 & 0.5353 & 29.50 & 0.5842 & 28.51 & 0.5783 & 29.47 & 0.5849 & 27.83 & 0.5443 & 29.48 & 0.5882 & \textbf{30.09} & \textbf{0.6186} \\
			S27 & 24.28 & 0.5935 & 26.61 & 0.6397 & 25.96 & 0.6556 & 26.88 & 0.6681 & 22.02 & 0.5639 & \textbf{26.97} & \textbf{0.6918} & 26.84 & 0.6371 \\
			S28 & 21.56 & 0.4588 & 23.74 & 0.4979 & 23.31 & 0.5017 & 23.83 & 0.5107 & 21.43 & 0.4221 & 23.75 & 0.5220 & \textbf{24.41} & \textbf{0.5243} \\
			S29 & 23.28 & 0.5826 & 25.00 & 0.6017 & 23.78 & 0.6009 & 25.06 & 0.6124 & 20.92 & 0.5152 & 24.84 & 0.6185 & \textbf{25.90} & \textbf{0.6304} \\
			S30 & 21.75 & 0.5534 & 23.38 & 0.5476 & 22.43 & 0.5588 & 23.56 & 0.5786 & 20.03 & 0.4332 & 23.18 & \textbf{0.5967} & \textbf{24.30} & 0.5926 \\
			S31 & 19.40 & 0.3890 & 21.33 & 0.4370 & 21.04 & 0.4417 & 21.36 & 0.4470 & 19.12 & 0.3099 & 21.08 & 0.4510 & \textbf{22.24} & \textbf{0.4584} \\
			S32 & 17.43 & 0.3227 & 19.41 & 0.3899 & 19.14 & 0.3936 & 19.40 & \textbf{0.3937} & 17.36 & 0.2288 & 18.94 & 0.3910 & \textbf{20.30} & 0.3936 \\
			S33 & 18.81 & 0.3378 & 21.36 & 0.4184 & 20.80 & 0.4208 & 21.27 & 0.4233 & 18.55 & 0.2783 & 21.07 & \textbf{0.4440} & \textbf{21.66} & 0.4141 \\
			S34 & 20.17 & 0.3673 & 21.57 & 0.4071 & 21.22 & 0.4078 & 21.63 & 0.4101 & 19.19 & 0.2743 & 21.33 & 0.4164 & \textbf{22.99} & \textbf{0.4377} \\
			S35 & 19.95 & 0.4260 & 22.82 & 0.4959 & 22.24 & 0.4960 & 22.78 & 0.4986 & 19.37 & 0.3468 & 22.36 & 0.4948 & \textbf{23.08} & \textbf{0.5070} \\
			S36 & 16.77 & 0.4675 & 19.43 & 0.5058 & 19.16 & 0.5261 & 19.54 & 0.5410 & 17.49 & 0.4392 & 19.29 & \textbf{0.5716} & \textbf{19.85} & 0.5051 \\
			S37 & 19.23 & 0.4341 & 21.03 & 0.4648 & 20.83 & 0.4747 & 21.17 & 0.4808 & 17.85 & 0.3264 & 20.94 & \textbf{0.4945} & \textbf{21.81} & 0.4846 \\
			S38 & 22.42 & 0.3352 & 25.09 & 0.4150 & 24.04 & 0.4062 & 25.02 & 0.4113 & 23.84 & 0.3111 & 24.84 & 0.4104 & \textbf{25.38} & \textbf{0.4291} \\
			S39 & 20.75 & 0.3768 & 22.99 & 0.4370 & 22.61 & 0.4380 & 23.05 & 0.4422 & 20.78 & 0.3211 & 22.89 & \textbf{0.4564} & \textbf{23.58} & 0.4508 \\
			S40 & 21.13 & 0.4392 & 23.25 & 0.4895 & 22.74 & 0.4899 & 23.29 & 0.4953 & 20.34 & 0.3720 & 23.00 & 0.5026 & \textbf{23.85} & \textbf{0.5032} \\
			S41 & 18.64 & 0.3863 & 21.01 & 0.4515 & 20.66 & 0.4576 & 21.04 & 0.4635 & 18.41 & 0.3082 & 20.65 & 0.4680 & \textbf{21.82} & \textbf{0.4986} \\
			S42 & 22.28 & 0.6301 & 25.11 & 0.6387 & 24.35 & 0.6414 & 25.22 & 0.6542 & 21.66 & 0.5609 & 24.94 & 0.6658 & \textbf{25.99} & \textbf{0.7007} \\
			S43 & 23.66 & 0.5652 & 25.45 & 0.6157 & 24.91 & 0.6214 & 25.59 & 0.6321 & 22.69 & 0.5330 & 25.50 & \textbf{0.6471} & \textbf{26.14} & 0.6021 \\
			S44 & 20.86 & 0.5394 & 23.89 & 0.6033 & 22.77 & 0.6068 & 24.00 & 0.6222 & 19.77 & 0.5142 & 23.88 & 0.6334 & \textbf{25.28} & \textbf{0.6502} \\
			S45 & 25.94 & 0.6612 & \textbf{28.46} & 0.7274 & 25.64 & 0.7199 & 28.04 & \textbf{0.7285} & 21.46 & 0.6002 & 27.62 & 0.7218 & 27.84 & 0.6904 \\
			\hline
			Avg & 21.49 & 0.4705 & 23.56 & 0.5190 & 22.84 & 0.5225 & 23.60 & 0.5301 & 20.56 & 0.4093 & 23.39 & 0.5387 & \textbf{24.54} & \textbf{0.5450} \\
			\hline
			\hline
		\end{tabular}
	\end{table*}

	We further compare the performance of our MA-GAN with bicubic interpolation, and some advanced SR methods including SRCNN \cite{dong2014srcnn}, FSRCNN \cite{dong2016fsrcnn}, VDSR \cite{kim2016vdsr}, EDSR \cite{lim2017edsr} and DBPN \cite{haris2018dbpn}, on the NWPU-RESISC45 \cite{cheng2017nwpu} dataset. SRCNN, FSRCNN and VDSR are the three CNN-based networks that minimize the $L_2$ loss while EDSR and DBPN minimize $L_1$ loss. All of these methods are optimized on the same dataset and environment for a fair comparison.
	
	In Section \ref{subsec:n_pcrdb_expe}, we reach the conclusion that there is little difference in the performance of the models when $N$ varies. Therefore, in this section, for each scale factor, we choose a better-performing MA-GAN. Specifically, for $r=2$, we choose MA-GAN(7); for $r=4$, we choose MA-GAN(3); and for $r=8$, we choose MA-GAN(4).
	
	Figures \ref{fig:result_x2}, \ref{fig:result_x4}, \ref{fig:result_x8} show the visual comparisons between different SR methods when the scale factor $r=2, 4, 8$, respectively. In these figures, we have selected only part of the $I_{SR}$ for magnified display. The selected image area marked with red boxes is rich in texture information and is difficult to recover. In this way, we can observe the performances of different methods. In our experiments, EDSR did not successfully learn the mapping of LR images to HR images, and the results were less than satisfactory. From the figure, we can also determine that although our method can achieve the relative best results, there is still a gap between the $I_{HR}$ and $I_{SR}$ generated by our model, mainly the difficulty of recovering those tiny textural features.
		
	Tables \ref{tab:results_x2}, \ref{tab:results_x4}, and \ref{tab:results_x8} show the quantitative comparisons when the scale factor $r=2, 4, 8$, respectively. We have compared the super-resolution performances of the different models for each of the image scenes in the NWPU-RESISC45 dataset to obtain a more detailed comparison. The highest PSNR values for each scene are marked in red, and the highest SSIM is marked in blue. 
	
	For $r=2$ (Table \ref{tab:results_x2}), MA-GAN(7) obtains an average PSNR of 31.98 dB and an average SSIM of 0.9102, both of which rank first among all the methods. The PSNR of MA-GAN(7) is 1.43 dB higher than the second-best value, while the SSIM is 0.0005 higher. In terms of PSNR, MA-GAN(7) outperforms almost all of the scenes and is ranked first in SSIM for nearly half of the scenes. With $r=4$ (Table \ref{tab:results_x4}), MA-GAN(3) performs best with an average PSNR of 27.62 dB and an average SSIM of 0.7640. The PSNR of MA-GAN(3) is 1.18 dB higher than the second-best value, while the SSIM is 0.0044 higher. The PSNR value of MA-GAN(3) is lower than those of the other comparison methods in 7 scenes. It obtains the highest SSIM value for 34 scenes and is lower than only DBPN in all other 11 scenes. The last one is the case of $r=8$ (Table \ref{tab:results_x8}), where MA-GAN(4) achieves the highest average PSNR of 24.54 dB and the average SSIM of 0.5450. The PSNR of MA-GAN(4) is 0.98 dB higher than the second-best result, while the SSIM is 0.0063 higher. MA-GAN(4) performs better in PSNR than almost all other methods and is not ranked first in only 4 scenes. With respect to SSIM, MA-GAM(7) performs better on 29 scenes.
	
	In summary, our MA-GAN outperforms all the other methods for $r=2,4,8$, respectively. In particular, MA-GAN is basically more than 1 dB higher than the second-best results among all methods in terms of PSNR values, with a maximum difference of 1.43 dB. However, the lead is not as large in SSIM. The result is only 0.0005 higher than DBPN at $r=2$. This is probably because the SR task of upsampling by a scale factor of 2 is simpler and the gap is not significant in this aspect. When $r$ increases, the advantage of MA-GAN in SSIM becomes more obvious, with a maximum improvement of 0.0063 over the second-place condition ($r=8$). Overall, although there are still some unsatisfactory aspects, our MA-GAN still performs the best.

	\section{Conclusion} \label{sec:conclusion}
	
	In this paper, we present a GAN-based SR network named the multi-attention-GAN that correctly learns the mapping from LR to HR images to generate perceptually pleasing HR images. Specifically, we first designed a GAN-based framework for the image SR task. The key to accomplishing the SR task is the image generator with post-upsampling that we designed. The main body of the generator contains two blocks; one is the PCRDB block, and the other is the AUP block. The AttPConv in the PCRDB block is a module that combines multi-scale convolution and channel attention to automatically learn and adjust the scaling of the residuals for better results. The AUP block is a module that combines pixel attention to perform arbitrary multiples of upsampling. These two blocks work together to help generate better quality images. For the loss function, we design a loss function based on pixel loss and introduce both adversarial loss and feature loss to guide the generator learning. Finally, it is demonstrated by our experiments that our proposed MA-GAN can perform better than some state-of-the-art SR methods.
	
	Our MA-GAN still encounters difficulty in generating tiny textures, which is a problem for all super-resolution algorithms, and we will continue to work on this issue in the future. Additionally, inspired by the satisfactory performance of our model, we will attempt to integrate our SR models into other vision tasks to help improve the results.
	
	\bibliographystyle{IEEEtran}
	\bibliography{IEEEabrv,henry}

\end{document}